# Bitcoin and its impact on the economy

Merrick Wang

**Abstract**

The purpose of this paper is to review the concept of cryptocurrencies in our economy. First, Bitcoin and alternative cryptocurrencies' histories are analyzed. We then study the implementation of Bitcoin in the airline and real estate industries. Our study finds that many Bitcoin companies partner with airlines in order to decrease processing times, to provide ease of access for spending in international airports, and to reduce fees on foreign exchanges for fuel expenses, maintenance, and flight operations. Bitcoin transactions have occurred in the real estate industry, but many businesses are concerned with Bitcoin's potential interference with the U.S. government and its high volatility. As Bitcoin's price has been growing rapidly, we assessed Bitcoin's real value; Bitcoin derives value from its scarcity, utility, and public trust. In the conclusion, we discuss Bitcoin's future and conclude that Bitcoin may change from a short-term profit investment to a more steady industry as we identify Bitcoin with the "greater fool theory," and as the number of available Bitcoins to be mined dwindles and technology becomes more expensive.

Bitcoin and its impact on the economy

**1. Introduction to Cryptocurrencies**

A cryptocurrency is a digital currency that uses cryptography for security. As a result of the cryptography, a cryptocurrency is difficult to counterfeit. The defining feature of a cryptocurrency is its organic nature: a cryptocurrency is not issued by a central authority, making it immune to governmental manipulation (Investopedia, n.d.).

*1.1 Definition of Money (Lee, 2009)*

Money is a medium of exchange that is accepted as a payment of goods or debts. Aristotle analyzed the problem of commensurability and concluded that money equalizes different things and makes it possible to compare otherwise unequal objects. His definition of money is as follows:
1. It must be durable.
2. It must be portable.
3. It must be divisible.
4. It must have intrinsic value.

Aristotle's definition of money has held true throughout history. For example, thousands of years ago, gold was a good form of currency.

*1.2 Definition of Money Applied to Cryptocurrencies*

The definition of money as detailed by Aristotle cannot be applied wholly to cryptocurrencies. Cryptocurrencies are not durable because they exist only virtually, but durability can be redefined as a currency's stability in existence. So far, cryptocurrencies have existed since 2009. Just like most other currencies around the world, cryptocurrencies are not backed by gold. Neither do cryptocurrencies have any intrinsic value. However, cryptocurrencies are portable and divisible on the realm of the Internet.

*1.3 Cryptocurrencies Advantages (Investopedia, n.d.)*
- The main advantage of the cryptocurrency is the avoidance of transfer fees typically involved in wire transfers or banks.
- Cryptocurrencies use a blockchain to store an online ledger of all past transactions. This limits the threat to hackers.
- As cryptocurrencies' payment processing is more efficient, it is used in crowdfunding in large companies such as JP Morgan.
- Cryptocurrency is easily accessible to the general public.

*1.4 Cryptocurrencies Drawbacks (Coinpupil, 2017)*
- Payments cannot be reversed.
- Cryptocurrencies are volatile in value since its prices are based on supply and demand.
- Lack of knowledge of cryptocurrencies in many companies and countries.

- If a cryptocurrency wallet is lost, there is no form of recovery, even with legal help.

## 2. Analysis of Bitcoin

*2.1 History*

Bitcoin was officially invented by Satoshi Nakamoto in 2008 and released to the public on January 8th, 2009. The creator's intentions of creating Bitcoin were to develop a cash-like payment system that not only allowed transactions to occur online but also have many of the benefits of physical cash (Berentsen & Schär, 2018).

*2.2 How Bitcoin Transactions Work (Kroeger, n.d.)*

A transaction for Bitcoin occurs when the public key of a Bitcoin is broadcasted to the Bitcoin network. Bitcoins are protected via a private key, a password that is unique to every Bitcoin. When both the private key and the public key are sent through a transaction to its recipient, the transaction is then authorized. A Bitcoin transaction cannot be reversed.

*2.3 Bitcoin's Price Chart*

Below is a summary of Bitcoin's price history:

| Date  | May 2010 | Feb. 2011 | Apr. 11, 2013 | Nov. 29, 2013 | Oct 21, 2017 | Dec. 15, 2017 | Oct 7, 2018 |
|-------|----------|-----------|---------------|---------------|--------------|---------------|-------------|
| Price | <$0.01   | $1.00     | $266          | $1,242        | $6,180       | $17,900       | $6535       |

*2.4 Analysis of Bitcoin's Price Chart*

Bitcoin was relatively unknown from the start, trading as a penny stock from 2009 to 2010. In February 2011, Bitcoin passed the $1 mark. Popularity for Bitcoin soared in 2014 due to China's interest in Bitcoin. In 2013, Jered Kenna, CEO of Bitcoin trading platform Tradehill, announced, "Bitcoin interest has risen tremendously in China over the last few months" (Hill, 2013). Additionally, Fred Ehrsam of Coinbase reported that trading activity in China for Bitcoin increased by ten times (Hill, 2013).

Another reason for Bitcoin's price increase in 2014 was due to excitement over the potential of Bitcoin. Many investors took the opportunity through an investment vehicle after venture capitalists identified Bitcoin as likely to grow (Hill, 2013).

In 2017, Bitcoin's price rapidly increased from $1,000 on January 1st, 2017, to $17,900 on December 15, 2017 (Rizzo, 2017). The 1,690% increase in price is attributed to rising interest in cryptocurrencies. Cryptocurrency fund manager Jacob Eliosoff says that "there is no question that these digital assets are experiencing growing interest, especially on the part of finance professionals and funds" (Bovaird, 2017). Since Bitcoin operates without a central agency, its popularity rises during periods of political turmoil.

A second reason for the enthusiasm over Bitcoin is the increase in initial coin offerings (ICOs). According to Pitchbook and Fortune, ICO's have raised more than $1

billion in the first 8 months of 2017. Pitchbook stated in July 2017, "ICOs had raised six times as much money in 2017 as they did during all of 2016" (Bovaird, 2017).

## 3. Analysis of Alternative Cryptocurrencies

Below is a look at the total market capitalization of cryptocurrencies as of October 8, 2018:

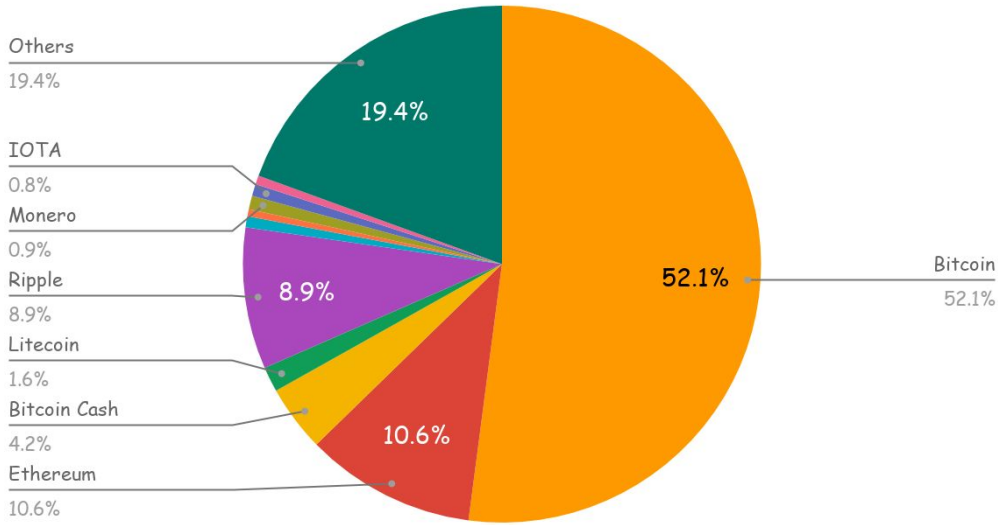

*Note.* Data from "Global Charts," n.d.

Cryptocurrencies other than Bitcoin often serve a specific purpose. Ethereum, for instance, is used specifically for smart contracts such as trade finance deals (Kharpal, 2017).  Bitcoin Cash transactions occur at faster rates than other cryptocurrencies (Kharpal, 2017). Ripple works with banks to eliminate percentage cuts with international payments or transfers (Kharpal, 2017).

## 4. Implementation

Many businesses have begun to accept Bitcoin as a method of payment. Most notably, Bitcoin has made its way into the airline and real estate industries.

*4.1 Airlines*

The Bitcoin company "Bitnet" formed a partnership with the Universal Air Travel Plan (UATP) in 2018 (Santos, 2018). UATP is a global payment platform managed by a group of airlines, which include Delta Air Lines and United Airlines. The CEO of UATP, Ralph A. Kaiser, said, "we are always looking for skilled partners to bring additional forms of payment to the UATP processing platform and Bitnet fits the bill well" (Santos, 2018). While the Bitnet partnership connects many major airlines to Bitcoin, Peach, a small Japanese airline, was the first to accept Bitcoin as payment for tickets in 2017.

Australia's Brisbane Airport has become the first in the world to accept Bitcoin at terminal shopping areas (Jenkins, 2018). Roel Hellemons, a manager at Brisbane Airport, commented, "many people around the world have made money investing in cryptocurrencies and a lot of these people travel internationally, so it makes sense to offer a digital currency experience within our terminals" (Jenkins, 2018).

Since Bitcoin is a digital currency independent of banks or central authorities, it can save airlines significant sums of money on foreign exchanges such as fuel expenses, maintenance, and flight operations. Many of these actions are performed overseas, and transactions may take long times to process. Bitcoin speeds up such transactions. Executive of Dubai-based Martin Consulting Mark Martin says, "There will be no more delays by bank holidays, weekends, or long festival breaks" (El Gazaar, 2017).

*4.2 Real Estate (Flemming & Molnar, 2018)*

During January 2017, a homebuyer used 3,300 bitcoins to buy a house for $3.225 million in Manhattan Beach. Real estate is perfectly suited for Bitcoin because "cryptocurrency is a way to send large amounts of money pretty easily with relatively low fees and little interference from middlemen," explains Neeraj Agrawal, director of communications at Coin Center.

While Bitcoin may seem a worthy alternative to cash, there seems to be no tax benefit when buying a house with Bitcoin. In fact, if the seller of the house decides to sell Bitcoin in exchange for cash, there are additional fees involved.

The realtor who facilitated the January 2017 transaction had difficulty working with Bitcoin as well. Justin Miller, who works with Beach City Brokers, had trouble finding a title and escrow company that would work with him on the deal. Many companies did not understand the concept of Bitcoin. Consequentially, Miller had to spend extra time finding a working title company for the transaction.

Another issue that occurs with Bitcoin users in the real estate industry is the proof of funds stage. Before an individual is allowed to purchase a home, he or she must first provide proof of funds to demonstrate that the loan can be paid off. The executive director of the Zcash Foundation, Josh Cincinnati, had difficulty with the homebuying process when he attempted to buy a house in Virginia. The loan officers that Cincinnati worked with had trouble with Bitcoin because it was not a "legitimate trade." The liquidity of Bitcoin, inherent in cryptocurrencies' nature, is met with significant ambiguity in the loan industry.

*4.3 Future of Implementation*

In the future, Bitcoin may be implemented by many more companies and industries as a form of payment. Bitcoin is wireless and allows transactions to occur quickly and easily. However, some concerns have been raised regarding Bitcoin's stability and legality. Many people still do not understand the concept of Bitcoin and other cryptocurrencies and may be unwilling to incorporate Bitcoin into their businesses.

Over the course of one day (between June 25, 2017 and June 26, 2017), Bitcoin's value decreased by over $200 (Williams, 2017). A business could conduct a transaction involving Bitcoin and lose significant sums of money if Bitcoin's value drops the following

day. Without a government to back and guarantee prices, Bitcoin will have a hard time being implemented in our economy.

Bitcoin may also draw the attention of regulators in the near future because it is not tied with a central agency. For instance, if Bitcoin owners decide to buy illegal drugs with Bitcoin, then the U.S. government may soon enact policies that will restrict the trading of Bitcoin as a currency.

Currently, Bitcoin is being used by many companies such as Microsoft, Paypal, Overstock, and DISH Network (Williams, 2017). In the near future, many other businesses may welcome Bitcoin with open arms.

**5. Bitcoin's Real Value**

While there are many incentives to using Bitcoin, one question still must be asked: "What is Bitcoin's worth, given the risks of investment?"

*5.1 Value of Bitcoin*

We want to assess Bitcoin's real value as its value rises. Based on the basic laws of supply and demand, Bitcoin has value because many people demand it. This value is Bitcoin's extrinsic value. We are interested in finding Bitcoin's intrinsic value. Former chairman of the Federal Reserve, Alan Greenspan, remarks, "The question is I do not understand where the backing of bitcoin is coming from… you have to really stretch your imagination to infer what the intrinsic value of bitcoin is. I haven't been able to do it" (Bloomberg, 2017).

Inherently, Bitcoin is becoming harder and harder to "mine," or create. It is more difficult to create more Bitcoins as the total number of existing Bitcoins nears its limit of 21 million Bitcoins. As a result, a scarcity of Bitcoins has been created, giving it some value to investors, in the same way that golds' scarcity gives it value.

Some value of Bitcoin also comes from its utility. Bitcoin is easily divisible, is not controlled by the government, and has an open source code, meaning the public can view and analyze transactions (Bearman, 2018). Bitcoin is constantly modified, slowly evolving as online services and transaction times improve.

The most fundamental element of Bitcoin's value, however, is trust. The American dollar can be paralleled to Bitcoin- while it costs around 16 cents to make a hundred dollar bill, the remaining $99.84 value comes from the trust that users of the dollar place around it. Bitcoin, in the same sense, is similar to a digital business; companies such as Facebook and Amazon are built on public trust, which gives them value. Bitcoin has value because the public believes in it and because it has the potential to eventually take the role of gold (a scarce resource) in our economies today (Bearman, 2018).

*5.2 Bitcoin Bubble: Greater Fool Theory*

In the 1982 paper "Bubbles, Rational Expectations, and Financial Markets," economists Olivier Blanchard and Mark Watson why Bitcoin is vulnerable to being in a bubble.  Bitcoin is similar to gold in that Bitcoin is not a company that will provide financial reports or pay out dividends. Individuals instead invest in Bitcoin for the same reasons that they invest in gold. People may want gold simply as a backup against inflation or economic collapse.

A second reason that people invest in gold is because they see the price of gold increase. As Blanchard and Watson put it, such an investor "bases his choice of whether or not to hold the asset on the basis of past actual returns rather than on the basis of market fundamentals" (Blanchard & Watson, 1982). In many instances, investors buy gold simply because the price is going up. They buy gold "not because of any fundamental economic insight or any analysis of value, but rather because they want to catch the train" (Thomson, 2017). These investors will hold onto gold as it appreciates and sell it to a "greater fool" before the value of gold declines, hoping that someone else values it even more than they do. Therefore, the "greater fool theory," an irrational belief, takes place with Bitcoin as it does with gold.

## 6. Bitcoin's Future

According to Credit Suisse's Global Markets Research Department, "the value of the cryptocurrency has been three times as volatile as the price of oil and 11 times more than the post-Brexit exchange rate between the dollar and the British pound" (Kindergan, 2017). Jeremy Grantham, who predicted the dot com bubble (2000) and the housing bubble (2007), speculates that Bitcoin is in a bubble and will crash soon (Cheng, 2018). The graph below displays how Bitcoin may be on the course of a bubble:

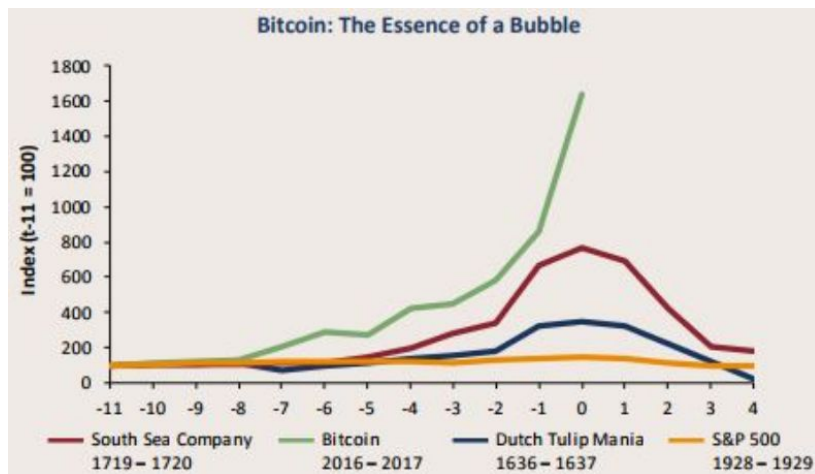

*Note.* Data from Cheng, 2018.

Ultimately, the government may interfere with Bitcoin in the near future. With government regulation, cryptocurrencies may become smaller and more plentiful, making the market more competitive. Bitcoin may change from a short-term profit investment to a more steady industry, especially as the number of available Bitcoins to be mined dwindles and technology becomes more expensive.

Bitcoin and its impact on the economy    8
## References

Bearman, S. (2018, January 16). As bitcoin's price plunges, skeptics say the cryptocurrency has no value. Here's one argument for why they're wrong. Retrieved from https://www.cnbc.com/2018/01/16/skeptics-say-bitcoin-has-no-value-heres-why-theyre-wrong.html

Berentsen, A., & Schär, F. (2018). *A Short Introduction to the World of Cryptocurrencies*. Retrieved from https://files.stlouisfed.org/files/htdocs/publications/review/2018/01/10/a-short-introduction-to-the-world-of-cryptocurrencies.pdf

Blanchard, O. J., & Watson, M. W. (1982, July). *NBER WORKING PAPER SERIES: BUBBLES, RATIONAL EXPECTATIONS AND FINANCIAL MARKETS* (Research Report No. 945). Retrieved from http://www.nber.org/papers/w0945.pdf

Bloomberg, J. (2017, June 26). What Is Bitcoin's Elusive Intrinsic Value? Retrieved from https://www.forbes.com/sites/jasonbloomberg/2017/06/26/what-is-bitcoins-elusive-intrinsic-value/#16cacce87194

Bovaird, C. (2017, September 1). Why Bitcoin Prices Have Risen More Than 400% This Year. Retrieved from https://www.forbes.com/sites/cbovaird/2017/09/01/why-bitcoin-prices-have-risen-more-than-400-this-year/#15d4160f6f68

Cheng, E. (2018, January 4). Investor who called last two major market crashes says bitcoin is a bubble. Retrieved from https://www.cnbc.com/2018/01/04/investor-who-called-two-market-crashes-says-bitcoin-is-a-bubble.html

Coinpupil. (2017, November 5). Advantages and Disadvantages of Cryptocurrency. Retrieved from https://coinpupil.com/altcoins/advantages-disadvantages-of-cryptocurrency/

El Gazaar, S. (2017, December 24). Can bitcoin fuel aviation around the world? Retrieved from https://www.thenational.ae/business/aviation/can-bitcoin-fuel-aviation-around-the-world-1.690130

Flemming, J., & Molnar, P. (2018, March 1). L.A.'s real estate industry enters the age of bitcoin. Retrieved from http://www.latimes.com/business/realestate/hot-property/la-fi-hp-bitcoin-real-estate-20180304-story.html

Global Charts. (n.d.). Retrieved from CoinMarketCap database.

Hill, K. (2013, October 23). Five Reasons For Bitcoin's Most Recent Price Surge. Retrieved from https://www.forbes.com/sites/kashmirhill/2013/10/23/five-possible-reasons-for-bitcoins-most-recent-surge/#ffa751b60ac1

Investopedia. (n.d.). Cryptocurrency. Retrieved from https://www.investopedia.com/terms/c/cryptocurrency.asp

Jenkins, A. (2018, January 30). This Major Airport Will Be the First in the World to Accept Bitcoin. Retrieved from https://www.travelandleisure.com/travel-news/brisbane-airport-bitcoin-cryptocurrency

Bitcoin and its impact on the economy    9Kharpal, A. (2017, December 14). Bitcoin vs. Ether vs. Litecoin vs. Ripple: Differences between cryptocurrencies. Retrieved from https://www.cnbc.com/2017/12/14/bitcoin-ether-litecoin-ripple-differences-between-cryptocurrencies.html

Kindergan, A. (2017, March 2). Is Bitcoin Safe? Retrieved from https://www.credit-suisse.com/corporate/en/articles/news-and-expertise/is-bitcoin-safe-201701.html

Kroeger, A. (n.d.). *Essays on Bitcoin* (T. Fuerst, Ed.). Retrieved from https://economics.nd.edu/assets/165129/alex_kroeger_essays_on_bitcoin.pdf

Lee, J. (2009, April 30). Aristotle and the Definition of Money. Retrieved from http://www.marketoracle.co.uk/Article10370.html

Rizzo, P. (2017, January 1). Bitcoin Price Tops $1,000 in First Day of 2017 Trading. Retrieved from https://www.coindesk.com/bitcoin-price-1000-january-1-2017/

Santos, M. (2018, January 2). 260 Airlines now accept Bitcoin thanks to new Bitnet Partnership. Retrieved from https://99bitcoins.com/260-airlines-accept-bitcoin-bitnet-partnership/

Thomson, D. (2017, December 9). Is Bitcoin the Most Obvious Bubble Ever? Retrieved from https://www.theatlantic.com/business/archive/2017/12/bitcoin-bubble/547952/

Williams, S. (2017, July 6). 5 Brand-Name Businesses That Currently Accept Bitcoin. Retrieved from https://www.fool.com/investing/2017/07/06/5-brand-name-businesses-that-currently-accept-bitc.aspx